\documentclass[12pt]{article}
\usepackage{a4}
\usepackage{latexsym} 
\usepackage{amssymb}  
\usepackage{amsmath} 
\usepackage{graphicx}
\usepackage{epsfig,wrapfig}
\usepackage[ usenames]{color}
\usepackage[rflt]{floatflt} 
\begin{document}

\newcommand{\talk}[3]
{\noindent{#1}\\ \mbox{}\ \ \ {\it #2} \dotfill {\pageref{#3}}\\[1.8mm]}
\newcommand{\stalk}[3]
{{#1} & {\it #2} & {\pageref{#3}}\\}
\newcommand{\snotalk}[3]
{{#1} & {\it #2} & {{#3}n.r.}\\}
\newcommand{\notalk}[3]
{\noindent{#1}\\ \mbox{}\ \ \ {\it #2} \hfill {{#3}n.r.}\\[1.8mm]}
\newcounter{zyxabstract}     
\newcounter{zyxrefers}        

\newcommand{\newabstract}
{\clearpage\stepcounter{zyxabstract}\setcounter{equation}{0}
\setcounter{footnote}{0}\setcounter{figure}{0}\setcounter{table}{0}}

\newcommand{\talkin}[2]{\newabstract\label{#2}\input{#1}}                

\newcommand{\rlabel}[1]{\label{zyx\arabic{zyxabstract}#1}}
\newcommand{\rref}[1]{\ref{zyx\arabic{zyxabstract}#1}}

\renewenvironment{thebibliography}[1] 
{\section*{References}\setcounter{zyxrefers}{0}
\begin{list}{ [\arabic{zyxrefers}]}{\usecounter{zyxrefers}}}
{\end{list}}
\newenvironment{thebibliographynotitle}[1] 
{\setcounter{zyxrefers}{0}
\begin{list}{ [\arabic{zyxrefers}]}
{\usecounter{zyxrefers}\setlength{\itemsep}{-2mm}}}
{\end{list}}

\newcommand{\citetwo}[2]{[\rref{y#1},\rref{y#2}]}
\newcommand{\citethree}[3]{[\rref{y#1},\rref{y#2},\rref{y#3}]}
\newcommand{\citefour}[4]{[\rref{y#1},\rref{y#2},\rref{y#3},\rref{y#4}]}
\newcommand{\citefive}[5]
{[\rref{y#1},\rref{y#2},\rref{y#3},\rref{y#4},\rref{y#5}]}
\newcommand{\citesix}[6]
{[\rref{y#1},\rref{y#2},\rref{y#3},\rref{y#4},\rref{y#5},\rref{y#6}]}

\begin{center}
{\large\bf Duality violations in hadronic $\tau$ decays and the value of $\alpha_s$}\\[0.5cm]
D.\ Boito,$^{a,b}$ O.\ Cat\`a,$^c$ M.\ Golterman,$^d$\footnote{Speaker at workshop}
M.\ Jamin,$^{e,b}$ K.\ Maltman,$^f$ J.\ Osborne,$^d$ S.\ Peris$^b$\\[0.3cm]
\null$^e$ICREA, \null$^a$IFAE, $^b$UAB, Barcelona, Spain, \null$^c$IFIC, Valencia, Spain,
\null$^d$SFSU, San Francisco, CA, USA, \null$^f$York Univ., York, Canada
\end{center}

Current estimates of the size of nonperturbative effects 
on the precision determination of $\alpha_s$ from hadronic $\tau$ decays
are incomplete for at least three reasons:
(i) In none of the existing determinations has the effect of duality violations (DVs) been
estimated quantitatively.  We demonstrated in previous work \cite{CGP,Manch} and in this
talk that effects from DVs may turn out to be numerically significant, and that the use
of only doubly-pinched moments of the $\tau$ spectral functions in order to suppress
DVs is not reliable.
(ii) While the moments used by ALEPH[4] and OPAL[5]
in principle probe the OPE up to dimension D=16, the analyses retain
contributions only up to D=8.
 This was shown to be not self-consistent \cite{MY}.
(iii) The most precise quoted values for $\alpha_s$ are based
on ALEPH data.  However, it is now known that correlations due to the unfolding of
spectral functions were omitted in ALEPH's latest data analysis \cite{Manch}.  Our preliminary
explorations indicate that the impact of this on the error in $\alpha_s$ is not necessarily
negligible.

The physical reason for the appearance of DVs is the presence of resonances in the
vector ($V$) and axial ($A$) $\tau$ spectral functions, which neither QCD perturbation theory nor the OPE describe adequately.  
As demonstrated in Ref.\ \cite{CGP}, it is possible that the pattern of
resonances is rather different in vector and axial channels, and this could lead to
a significant effect on the value of $\alpha_s$ even if the sum of $V$ and $A$ 
spectral functions ($V+A$) looks relatively flat in the region between $s\approx 1$\ GeV$^2$ and
$s=m_\tau^2$.   In other words, the assumption  that DVs can be
safely neglected in $V+A$ may turn out not to be justified.   Because of this and point (ii)
above, the estimates of the nonperturbative part of $R_\tau$ (and other moments)
used in the literature need to be revisited.

\begin{figure}[tb]
\vspace*{0ex}
\begin{center}
\includegraphics*[width=12cm,height=7cm]{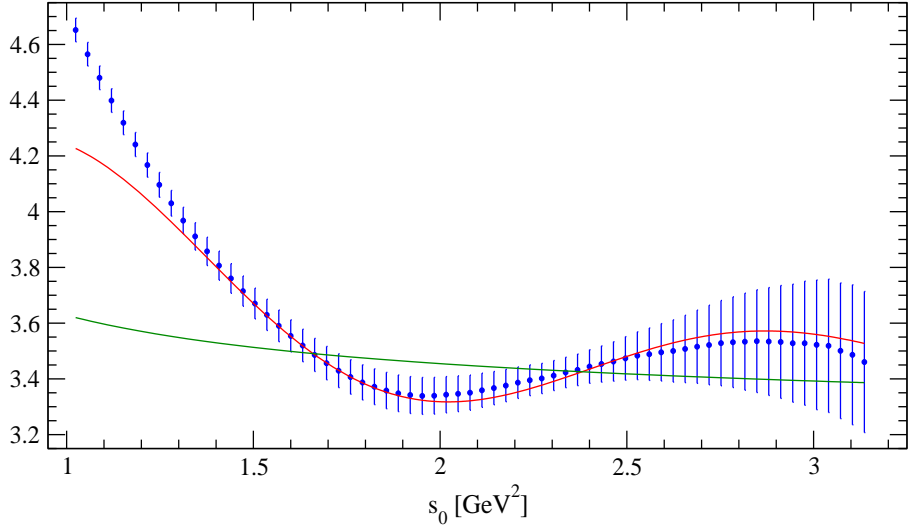}
\caption{{\it Fit to Eq.~(\ref{w1}), with the ansatz~(\ref{ansatz}) (red curve) and
without the ansatz (green curve).
}}
\label{fig1}
\end{center}
\vspace{-0.3cm}
\end{figure}

There is no quantitative theory of DVs in QCD, and in order to investigate this issue
in more detail, one has to resort to a (physically reasonable) model \cite{CGP}.  
This makes it possible to explore finite-energy sum rules with simple,
nonpinched weights.   Figure~\ref{fig1} shows a fit to the nonstrange integrated
vector spectral function using the sum rule
\begin{equation}
\label{w1}
\int_0^{s_0}ds\;\rho_{exp}(s)=-\frac{1}{2\pi i}\oint_{|s|=s_0} ds\;\Pi_{OPE}(s)
-\int_{s_0}^\infty ds\;\rho_{DV}(s)\ .
\end{equation}
Here $\rho_{exp}(s)$ is the experimental spectral function,
available up to $s_0=m_\tau^2$, $\Pi_{OPE}(s)$  the OPE expression for the
vacuum polarization, and $\rho_{DV}(s)$ the ``duality-violating'' part of the 
spectral function, {\it i.e.}, the part not present in $\mbox{Im}\;\Pi_{OPE}(s+i\epsilon)$.
The contour integral over  $\Pi_{OPE}(s)$ is parametrized by $\alpha_s(m_\tau^2)$ and the OPE condensates.   The functional form of $\rho_{DV}(s)$ is not known from first principles.  Following Ref.~\cite{CGP}, we model it as
\begin{equation}
\label{ansatz}
\rho_{DV}(s)=\kappa\;e^{-\gamma s}\;\sin{(\alpha+\beta s)}\ ,
\end{equation}
which introduces four new parameters into the fits to experimental data.   We restrict
$s_0$ to an interval $s_0\in[s_{min},m_\tau^2]$, 
and we vary $s_{min}$ in order to check for stability.  For the 
physical motivation of the model of Eq.~(\ref{ansatz}), see Ref.~\cite{CGP}.
The fit in Fig.~\ref{fig1} was done with OPAL data \cite{OPAL}, and $s_{min}=1.5$~GeV$^2$.  The result for
$\alpha_s(m_\tau^2)$ is
\begin{equation}
\label{alphas}
\alpha_s(m_\tau^2)=0.307(18)(4)(5)\ \quad(\mbox{FOPT})\ ,\qquad
\alpha_s(m_\tau^2)=0.322(25)(7)(4)\ \quad(\mbox{CIPT})\ ,
\end{equation}
where errors are the $\chi^2$ error from the fit, from varying $s_{min}=1.5\pm 0.1$~GeV$^2$,
and from varying the estimated coefficient of $(\alpha_s/\pi)^5$ in the Adler function in the range
$0$ to $566$.

Fig. 1 demonstrates that DVs cannot, in general, be neglected in
analyzing hadronic $\tau$ decay data, even integrated versions thereof. This
is confirmed by explorations which also include the axial channel in the
fits. Essentially all existing $\tau$-based extractions of $\alpha_s$ neglect
DVs, and thus assume model (2) with $\kappa$ set
to zero by hand. Such an assumption is necessarily accompanied by an
uncertainty not accounted for in currently quoted errors for $\alpha_s$.
Fig. 1 shows that it is dangerous to assume, {\it a priori}, that this
uncertainty can be neglected for analyses involving doubly-pinched
weights.  We are presently studying different combinations of $V$ and
$A$ spectral-function moments with the goal of quantifying this
uncertainty.  This may help reducing the fitting errors on $\alpha_s$, but at
present, if one wishes to be conservative, and stick only to results not
subject
to additional, currently unquantified uncertainties, one should
treat $\alpha_s$ from $\tau$ decays as being known to at best the accuracy given in
Eq.~(\ref{alphas}).

%

\end{document}